\documentclass[aip,jcp,floatfix,citeautoscript,preprint]{revtex4-1}
\setcitestyle{super}

\usepackage{graphicx}
\usepackage{setspace}[1.5] 
\usepackage{amsmath, amsthm, amssymb}
\usepackage{xfrac}
\usepackage{dcolumn}
\usepackage{gensymb}
\usepackage{color}

\begin{document}

\date{\today}

\title{Water on hexagonal boron nitride from diffusion Monte Carlo}

\author{Yasmine S. Al-Hamdani}
\affiliation{Thomas Young Centre and London Centre for Nanotechnology,
  17--19 Gordon Street, London, WC1H 0AH, U.K.}
\affiliation{Department of Chemistry, University College London, 20
  Gordon Street, London, WC1H 0AJ, U.K.}

\author{Ming Ma}
\affiliation{Thomas Young Centre and London Centre for Nanotechnology,
  17--19 Gordon Street, London, WC1H 0AH, U.K.}
\affiliation{Department of Chemistry, University College London, 20
  Gordon Street, London, WC1H 0AJ, U.K.}

\author{Dario Alf\`{e}}
\affiliation{Thomas Young Centre and London Centre for Nanotechnology,
  17--19 Gordon Street, London, WC1H 0AH, U.K.}
\affiliation{Department of Earth Sciences, University College London,
  Gower Street, London WC1E 6BT, U.K.}

\author{O. Anatole von Lilienfeld} \affiliation{Institute of Physical
  Chemistry and National Center for Computational Design and Discovery
  of Novel Materials, Department of Chemistry, University of Basel,
  Klingelbergstrasse 80 CH-4056 Basel, Switzerland}
\affiliation{Argonne Leadership Computing Facility, Argonne National
  Laboratories, 9700 S. Cass Avenue Argonne, Lemont, Illinois 60439, USA}

\author{Angelos Michaelides}
\email{angelos.michaelides@ucl.ac.uk}
\affiliation{Thomas Young Centre and London Centre for Nanotechnology,
  17--19 Gordon Street, London, WC1H 0AH, U.K.}
\affiliation{Department of Chemistry, University College London, 20
  Gordon Street, London, WC1H 0AJ, U.K.}

\begin{abstract}
Despite a recent flurry of experimental and simulation studies, an
accurate estimate of the interaction strength of water molecules with
hexagonal boron nitride is lacking. Here we report quantum Monte Carlo
results for the adsorption of a water monomer on a periodic hexagonal
boron nitride sheet, which yield a water monomer interaction energy of
$-84\pm5$ meV. We use the results to evaluate the performance of
several widely used density functional theory (DFT) exchange
correlation functionals, and find that they all deviate
substantially. Differences in interaction energies between different
adsorption sites are however better reproduced by DFT.
\end{abstract}
\maketitle 

Hexagonal boron nitride (h-BN) has become popular for anyone with an
interest in 2-dimensional materials, due to a number of notable
properties such as high thermal conductivity, mechanical robustness
and exceptional resistance to oxidation\cite{pakdel_nano_2014}, and
not least because it is isostructural with graphene.  Our interest has
been piqued by experimental reports of fascinating behavior of water
at h-BN such as superhydrophobicity\cite{bn_exp3}, water cleaning
ability\cite{Lei_13} or generation of electric
current\cite{Siria_13}. These experiments have already prompted a
number of simulation studies of water on h-BN sheets and
nanostructures using both density functional theory (DFT) and
classical molecular
dynamics\cite{Tocci_2014,bn_exp,li_wetting_2012,won_water_2007,won_structure_2008,gordillo_wetting_2011,dutta_wetting_2011}. They
have been incredibly informative and have helped to e.g. understand
the electrical currents generated in BN nanotubes\cite{Siria_13}.

However, there is one major unknown at the very heart of any water/BN
simulation study: we simply do not know what the interaction strength
of a water molecule with h-BN is. DFT calculations yield a range of
values for the water monomer adsorption energy depending on the
exchange-correlation (xc) functional
used\cite{Tocci_2014,bn_exp,li_wetting_2012} and force fields rely on
interaction parameters fitted to particular xc functionals or to
experimental data such as contact angles for macroscopic water
droplets\cite{won_structure_2008,won_water_2007,gordillo_wetting_2011,dutta_wetting_2011}. If
fitting to experiment one needs to be certain that the experimental
conditions are exactly known; recent lessons learned for water
droplets on graphene reveal that contact angle measurements are
incredibly sensitive to surface preparation conditions and levels of
cleanliness\cite{taherian_what_2013,rafiee_wetting_2012,shih_breakdown_2012,li2013effect}.

The lack of well-defined reference data for water on h-BN is
representative of a much broader problem: there are very few systems
for which accurate water monomer adsorption energies have been
established. Mainly this is because even at low temperatures water
molecules cluster into larger aggregates making the determination of
monomer adsorption energies with established surface science
techniques such as temperature programmed desorption or single crystal
adsorption calorimetry highly
challenging\cite{hodgson_water_2009,campbell_enthalpies_2013,carrasco_molecular_2012}.
In the absence of experimental data simulations play an important
role, either via explicitly correlated quantum chemistry approaches or
quantum Monte Carlo (QMC) (see \textit{e.g.}  references
\citenum{Li2008L135,Ma_11b,Ma_11a,Voloshina_11,jenness2010benchmark,shulenburger_quantum_2013,lu_2009}). Indeed
given recent increases in computational capacity and the fact that it
can be applied to periodic systems, QMC has emerged as a powerful
technique for obtaining interaction energies of molecules with
surfaces\cite{Ma_11b,Ma_11a} or
biomolecules\cite{Mitas,benali_application_2014,al-hamdani1}.

Here we report results for interaction energy curves for water on a
periodic h-BN sheet using fixed node diffusion Monte Carlo (DMC). We
obtain from this an estimate of the water/h-BN interaction strength of
about $-84\pm5$ meV at an equilibrium water-surface distance of
\textit{ca.} 3.4 {\AA}. We have also computed interaction energy
curves with a variety of DFT xc functionals and we find that these
differ significantly from DMC. Except for LDA, of the functionals
considered those that do not account for van der Waals underbind and
those that do, overbind. DFT based predictions of the equilibrium
adsorption height are much better with several functionals coming
within 0.2 {\AA} of DMC. In addition, based on DMC and DFT
calculations of water on h-BN in a second metastable adsorption
structure, we find that many of the xc functionals do reasonably well
in predicting the relative energy difference between the stable and
metastable adsorption structures. 

Two different levels of theory have been used in this study, fixed
node DMC and DFT. A standard computational setup has been used for
each and so we only discuss the key features here.
\begin{figure}
\centering \includegraphics[width=0.45\textwidth]{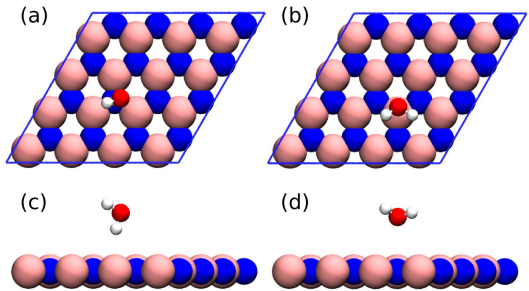}
\caption{Structures of the two adsorption modes of water on h-BN
  considered in this study. (a,c) Top and side view of water above an
  N site of h-BN. (b,d) Top and side view of water above a B site of
  h-BN. Boron is pink, nitrogen is blue, oxygen is red and hydrogen is
  white. All calculations have been performed on periodic unit cells,
  with the periodic unit cell in the x,y plane indicated by the blue
  frames in (a) and (b).}\label{configs}
\end{figure}

QMC calculations were undertaken using the CASINO code\cite{casino},
with Slater-Jastrow type trial wavefunctions in which the Jastrow
factor contains electron-nucleus, electron-electron, and
electron-electron-nucleus terms. We used Trail and Needs
pseudopotentials\cite{TN1,TN2} for all atoms, in which the 1s
electrons of B, N, and O were treated as core.  This set-up for the
DMC calculations is similar to the one used in reference
\citenum{al-hamdani1} where water adsorption was examined on
1,2-azaborine and agreement between DMC and coupled cluster with
single, double and perturbative triple excitations (CCSD(T)) to within
9 meV was obtained. The initial single particle wavefunctions for use
in DMC were obtained from DFT plane-wave calculations using the PWSCF
package\cite{pwscf}. A standard 300 Ry energy cut-off was applied and
for efficiency the resulting wavefunctions were expanded in terms of
B-splines \cite{bsplines} using a grid multiplicity of 2.0. Trial
wavefunctions were generated using the local density approximation
(LDA)\cite{LDA} which has been validated for weak interactions in
previous work\cite{alfe1,al-hamdani1}. After optimization of the trial
wavefunctions in variational Monte Carlo, we used 6,553,840 walkers
across 16,384 cores for each point along the DMC interaction energy
curves. The locality approximation was utilized\cite{locapp} with a
time step of 0.015 a.u. which we tested against a time step of 0.005
a.u.

VASP 5.3.5\cite{vasp1,vasp2,vasp3,vasp4} was used for the DFT
calculations, making use of projector augmented wave (PAW)
potentials\cite{PAW_94,PAW_99} to model the core regions of atoms
(again the 1s electrons of B, N, and O were treated as core).
Following careful tests, we chose a 500 eV plane-wave cut-off and a
($4\times4$) unit cell of h-BN with 16 {\AA} between sheets, along
with $\Gamma$-point sampling of reciprocal space.\footnote{For
  example, tests with a higher plane-wave cut-off (600 eV) and denser
  \textbf{k}-point mesh ($5\times5\times1$) performed for the PBE
  functional yielded an interaction energy that differed from the
  reported one by $<3$ meV.  Similarly when we tested the current
  set-up for water adsorption on the 1,2-azaborine system considered
  in reference \citenum{al-hamdani1} against all-electron PBE
  calculations with an aug-ccPV5Z basis set we found that the results
  with the two approaches differed by only 1 meV; the PBE adsorption
  energy for that system being 109-110 meV.}  The proliferation of DFT
xc functionals over the last decade\cite{burke2012perspective} means
that we cannot possibly consider all xc functionals or even all modern
xc functionals designed to capture weak interactions. Rather we
consider a small selection that have been widely used in adsorption
studies. This includes the LDA, the PBE\cite{PBE} generalised gradient
approximation (GGA), two hybrid functionals (PBE0\cite{PBE0a,PBE0b}
and B3LYP\cite{b3lypA,b3lypB,b3lypC,b3lypD}), and several van der
Waals (vdW) inclusive functionals (PBE+D2\cite{D2},
PBE+D3\cite{D3a,D3b}, DFT+vdW\cite{SCS},
optB86b-vdW\cite{vdwDF,B86,vdw_opt11} and vdW-DF2\cite{vdwDF,vdwDF2}).
The DFT+vdW correction (from Tkatchenko and Scheffler) was applied to
three xc functionals (PBE, PBE0 and B3LYP).
\begin{figure}
\centering \includegraphics[width=0.5\textwidth]{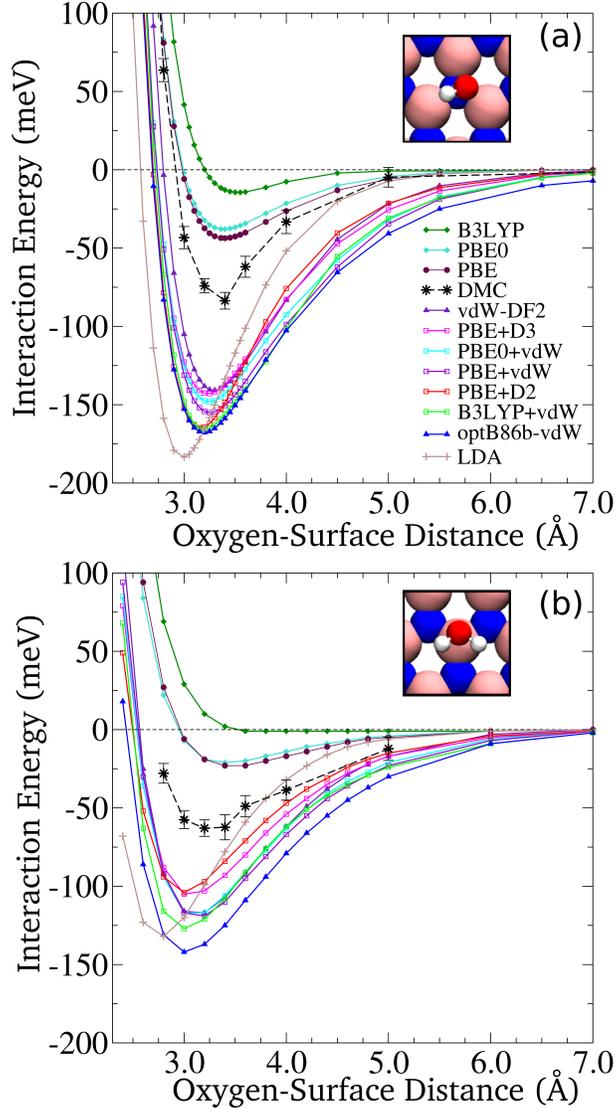}
\caption{(a) Interaction energy curves for water situated above the N
  site in h-BN as shown in the inset. (b) Interaction energy curves
  for water situated above the B site in h-BN as shown in the
  inset. The lines connecting the data points are merely there to
  guide the eye.}\label{BC-fig}
\end{figure}

Results for interaction energy curves of a water monomer on h-BN in
two different adsorption modes (Figure \ref{configs}), obtained from
DMC and a range of DFT xc functionals, are shown in Figure
\ref{BC-fig}. The interaction energy between the adsorbate and
substrate is plotted as a function of the perpendicular distance
between the oxygen atom of the water molecule and the h-BN sheet. The
absolute interaction energy between water and the substrate were
calculated as follows,
\begin{equation}
E_{int}=E^{tot}_{d}-E^{tot}_{\text{far}}\label{BE}
\end{equation}
where $E^{tot}_{d}$ is the total energy of water and h-BN at a given
oxygen-surface separation distance, d, and $E^{tot}_{\text{far}}$ is
the total energy of water and h-BN at 8 {\AA} oxygen-surface
distance. This definition allows the same Jastrow factor to be used
for all configurations, including the reference structure in DMC.
Adsorption structures were obtained from optB86b-vdW optimizations of
water on a fixed h-BN sheet, see Supporting Information for
coordinates of the adsorption and reference
configurations\cite{SI_ref}.  We chose to use the same structures for
DMC and all xc functionals because this makes for a cleaner
comparison. For many of the xc functionals we have computed
interaction energy curves with fully relaxed structures and the
differences between the relaxed and the fixed structures are $<5$ meV,
except in the repulsive wall at short oxygen-surface separations. The
first adsorption structure considered has the oxygen of the water
molecule above an N site with one of the OH bonds directed at that N
atom (Figure \ref{configs}(a,c)). This is the most stable adsorption
structure according to previous DFT studies \cite{Tocci_2014}. The
second structure has the oxygen atom of the water molecule above a B
site with the plane of the molecule tilted away from the substrate by
128{\degree} (Figure \ref{configs}(b,d)). According to our DFT
calculations this is the most stable structure for water at the B site
but $\sim20$ meV less stable than the N site adsorption structure. We
consider water adsorption at the B site to establish if the DFT site
preference for this system is correct\cite{Feibelman_2001}.

Let us now focus on the DMC interaction energy curves for water on
h-BN. Because of the enormous computational cost of DMC we can only
compute a small number of points for each energy curve, which limits
the resolution of the curves. Nonetheless they are sufficiently well
defined to yield an adsorption energy of $-84\pm5$ meV at a height of
$\sim3.4$ {\AA} at the N site and an adsorption energy of $-63\pm5$ at
a height of $\sim3.2$ {\AA} at the B site. The O atom sits slightly
further away from the substrate at the N site because of the
orientation of the molecule at this site, wherein the H atoms points
to the N, forming a weak hydrogen bond like interaction. The relative
energies of the two sites confirms the DFT site preference but more
importantly provides an estimate of the water monomer interaction
energy that is free of any arbitrary choices of DFT xc
functional.

Obtaining an accurate estimate of the interaction strength between
water and h-BN is important in its own right, however, it also
provides a valuable benchmark which we now exploit. Here, we use our
DMC derived interaction energy curves to evaluate how various DFT xc
functionals perform for this system.  Interaction energy curves from
several functionals are included in Figure \ref{BC-fig} and in some
respects these reveal a familiar story. Looking at the most stable
site first, LDA overbinds by predicting an adsorption energy of $-183$
meV with the molecule $0.4$ {\AA} closer to the surface than DMC. In
contrast the GGA and the hybrid functionals underbind: PBE is
$\sim-45$ meV, PBE0 is $\sim-40$ meV and B3LYP is $\sim-15$ meV. The
PBE and PBE0 adsorption heights are fairly reliable at $3.40$ {\AA}
whereas the shallow B3LYP minimum is located at $3.55$ {\AA}. More
interesting are the results from the vdW inclusive functionals since
these are in principle designed to accurately describe weak
interaction systems. Surprisingly, we find that all vdW inclusive
functionals considered significantly overbind this adsorption
system. Specifically the adsorption energies are in the $-140$ to
$-170$ meV range, with vdW-DF2 predicting the smallest adsorption
energy and optB86b-vdW the largest. This overbinding also persists at
large adsorbate-substrate distances; compare for example the DMC and
vdW-inclusive DFT results at $4-5$ {\AA} from the surface. The
predicted height above the surface is in reasonably good agreement
with DMC, only around $0.1$ {\AA} closer to the surface for all
vdW-inclusive functionals.

Moving to the B site adsorption structure we find that systematically,
with the exception of PBE+D2, the interaction strength is reduced by
$\sim20-30$ meV. This is in very good agreement with the DMC energy
difference between these two sites (PBE-D2 predicts that the B site is
$\sim60$ meV less stable than the N site). Thus although none of the
xc functionals considered come within $40\%$ of the DMC interaction
energy, the change in interaction energies between adsorption sites
are in most cases described fairly accurately. We note that we have
considered just two adsorption structures and considerably more work
is needed to fully substantiate this conclusion. Moreover, at this
stage we do not fully understand the poor performance of the vdW
functionals and defer a more detailed analysis to a future publication
in which results from on-going water adsorption calculations on BN
clusters will also be presented.

In summary, we have obtained DMC interaction energy curves for water
on a periodic hexagonal sheet of BN and used these to evaluate the
performance of a number of xc functionals. The interaction energy
obtained is $-84\pm5$ meV. This is clearly a small number;
corresponding to the physisorption regime.\footnote{Zero point energy
  contributions (computed within the harmonic approximation) weaken
  the optB86b-vdW interaction strength by $\sim30$ meV. Since this is
  the most strongly binding xc functional, others are likely to show a
  smaller reduction than this.} It is, however, about 15 meV larger
than the value predicted by DMC for water on graphene. Interestingly
many of the van der Waals inclusive functionals also predict a similar
15--20 meV increase upon going from graphene to
h-BN\cite{Tocci_2014,Ma_11b}. This suggests that although interaction
energies are overestimated with these functionals, the relative
interaction energies between the two materials are fairly well
described.

We are grateful for support from University College London and Argonne
National Laboratory (ANL) through the Thomas Young Centre-ANL
initiative. Some of the research leading to these results has received
funding from the European Research Council under the European Union's
Seventh Framework Programme (FP/2007-2013) / ERC Grant Agreement
number 616121 (HeteroIce project). A.M. is supported by the Royal
Society through a Wolfson Research Merit Award. O.A.v.L. acknowledges
funding from the Swiss National Science foundation (No. PP00P2
138932). This research used resources as part of an INCITE project
(awarded to D.A.)  at the Oak Ridge National Laboratory (Titan), which
is supported by the Office of Science of the U.S. Department of Energy
(DOE) under Contract No. DEAC05-00OR22725. This research also used
resources of the Argonne Leadership Computing Facility at Argonne
National Laboratory, which is supported by the Office of Science of
the U.S. DOE under contract DE-AC02-06CH11357. In addition, we are
grateful for computing resources provided by the London Centre for
Nanotechnology and University College London.

\end{document}